# Turning Insulators into Accelerators: Deciphering the Interfacial Conductivity Boost in ZrO$_2$-Li$_2$ZrCl$_6$ Composites through Machine Learning Molecular Dynamics Simulations

Boyuan Xu, Chen Qian, Liyi Bai, Chenlu Wang, Feng Ding, Qisheng Wu[*]

Suzhou Laboratory, Suzhou, Jiangsu 215126, People's Republic of China

[*]Corresponding author: wuqs@szlab.ac.cn

**Abstract**

Halide solid-state electrolytes have emerged as promising candidates for all-solid-state lithium batteries due to their high oxidative stability and deformability, yet their moderate ionic conductivity remains a bottleneck. While incorporating ionically insulating ZrO$_2$ nanoparticles (*Nat. Commun.* 2023, 14, 2459) has been experimentally shown to enhance the ionic conductivity of Li$_2$ZrCl$_6$, the atomistic origin governing this interfacial phenomenon remains unclear. Here, we bridge the spatiotemporal gap in modeling complex heterostructures by developing an accurate machine-learned force fields based on neuroevolution potential, enabling large-scale molecular dynamics simulations of ZrO$_2$/Li$_2$ZrCl$_6$ heterostructures. By systematically investigating four representative low-lattice-mismatch ZrO$_2$/Li$_2$ZrCl$_6$ interfaces, we identify spontaneous interfacial amorphization driven by space-charge effects upon surface cleavage, trapping Li$^+$ and leading to under-coordinated Li$^+$ polyhedrons with pronounced geometric distortion. These distorted amorphous interfacial regions exhibit markedly enhanced Li$^+$ hopping activity, significantly outperforming the bulk lattice, provided that local mobile Li$^+$ inventory is not depleted by surface charge redistribution. This work establishes a computational framework for training validated machine-learned force fields for interfaces and provides mechanistic understandings of the interfacial conductivity boost in the insulator-conductor composites, guiding the rational design of electrolytes toward next-generation solid-state batteries.

# 1. Introduction

All-solid-state lithium batteries (ASSLBs) have emerged as the premier successors to conventional liquid-electrolyte-based lithium-ion batteries (LIBs) in many applications such as electric vehicles, offering enhanced safety profiles, superior thermal stability, and the potential for higher energy densities.[1-3] The viability of ASSLBs hinges on the solid-state electrolyte (SSE), a pivotal component that must simultaneously satisfy stringent performance requirements including high ionic conductivity (typically >1 mS/cm),[4] adequate mechanical compliance,[5] and a wide electrochemical stability window (typically >4 V)[6]. Within the landscape of inorganic SSEs, sulfide electrolytes demonstrate exceptional ionic conductivities comparable to liquid organic electrolytes,[3] however, their practical application is impeded by the severe moisture sensitivity and poor compatibility with cathode materials.[7-10] Conversely, while oxide SSEs exhibit superior chemical and electrochemical stability, their widespread deployment is constrained by inherently low ionic conductivities and poor mechanical processability.[11-14]

Halide SSEs had been largely overlooked until 2018, when Asano *et al.* reported the $Li_3YCl_6$ and $Li_3YBr_6$ with ionic conductivity of >1 mS/cm at room temperature and wide electrochemical stability windows (~4 V),[15] which has boosted a surge of research into isostructural chlorides, specifically the $Li_3MCl_6$ (M=Sc, In, Er) type SSEs[16-21]. While these materials offer a compelling balance of high ionic conductivity and oxidative stability,[21, 22] their commercial viability is severely constrained by the prohibitive cost and the scarcity of the requisite rare-earth or indium precursors.

Within this landscape, $Li_2ZrCl_6$ (LZC) stands out as a cost-effective alternative, retaining the favorable mechanical and chemical attributes of the halide family.[23] However, its practical utility is currently limited by suboptimal room-temperature ionic conductivity.[24] Experimental reports indicate a wide conductivity variance (0.001~1.0 mS/cm) for $Li_2ZrCl_6$, being highly sensitive to synthesis protocols.[23, 25-28] While high conductivities (0.77~1.0 mS/cm, activation energy $E_a$ =0.33~0.38 eV) have been achieved in ideal- and Zr-deficient phases, these results necessitate intricate preparation methods, such as two-step mechanochemical synthesis[27] or prolonged ball-milling with intermittent manual homogenization[29]. In contrast, the $\alpha$-$Li_2ZrCl_6$ phase, prepared via straightforward one-step ball-milling, typically delivers only moderate conductivities (0.29~0.81 mS/cm, $E_a$=0.35 eV). Furthermore, thermal annealing induces a phase transition to the $\beta$-$Li_2ZrCl_6$ polymorph, which suffers from a detrimental two-order-of-magnitude reduction in conductivity and a significantly increased $E_a$ (0.50 eV).[23, 24, 26, 28]

Apart from the aliovalent doping strategy (e.g., the Zr site substitution with Y, Er, Sc)[30-32], the design of heterostructure systems leveraging interfacial conduction enhancement has emerged as a promising alternative. Anomalous ionic transport enhancement has been documented in various composite architectures, such as $CaF_2$/$BaF_2$ multilayered film,[33] LiF/silica films,[34] and polymer electrolytes with inorganic fillers,[35] though the fundamental mechanisms governing this interfacial boost in conductivity remain incompletely understood. Recently, Kwak *et al.* reported a counter-intuitive observation that the incorporation of ionically insulating $ZrO_2$ nanoparticles into $Li_2ZrCl_6$ resulted in a 50% to 200% increase in $Li^+$ conductivity, contingent on the $ZrO_2$ grain sizes.[36] Despite these encouraging experimental results, the atomic-scale origins of this interfacial conductivity boost has not been fully resolved. Herein, a large-scale and long-time simulation with high accuracy was desperately needed to unveil the dynamical interface ionic conduction mechanism that is beyond the capability of direct experimental observations.

In this work, we present a comprehensive computational investigation into the atomistic evolution of $ZrO_2$/$Li_2ZrCl_6$ interfaces and its impact on $Li^+$ diffusion dynamics, revealing the microscopic origins of the interfacial conductivity enhancement in the halide composite electrolytes. Recognizing that local atomic environments, particularly $Li^+$ coordination geometry and polyhedral distortions, are intimately linked to the ion mobility, we correlate the spatial distribution of $Li^+$ hopping events with interfacial structural features to probe how local disorder modulates transport. To bridge the gap between *ab initio* accuracy and the spatiotemporal scales required for realistic interface modeling, we developed a robust machine learning force field (MLFF) within the neuroevolution potential (NEP) framework through a computationally efficient active learning workflow. Enabled by this MLFF, we performed large-scale (~$10^4$ atoms), long-timescale (20 ns) machine learning molecular dynamics (MLMD) simulations of four $ZrO_2$/$Li_2ZrCl_6$ interfaces with lower than 1% lattice mismatch. Our simulations reveal that the formation of the space-charge layer upon surface cleavage drives a spontaneous structural reconstruction at the interfacial region. This triggers local amorphization, generating a high concentration of undercoordinated and geometrically distorted $Li^+$ polyhedrons. We demonstrate that while a fraction of these disordered units is immobilized to satisfy surface charge neutrality, the excess mobile $Li^+$ within the amorphous domains facilitate rapid ionic transport. These findings provide critical atomistic insights into the interplay between interfacial disorder, space-charge effects, and defect engineering, offering new design principles for high-performance composite SSEs.

## 2. Results and discussions

### 2.1. Development and validation of the MLFF

A high-fidelity NEP model was developed to reproduce the potential energy surface (PES) derived from (DFT) calculations, through a rigorous active learning scheme, as illustrated in **Figure 1a**. To expedite convergence and mitigate initialization bias, the training set was seeded by coupling the MLMD simulations with the MACE-MPA-0[37] foundation model, utilizing a SOAP-based descriptor for structural selection. This strategy significantly enhanced sampling efficiency and minimized the requisite active learning iterations. Subsequently, an active learning approach was adopted, wherein four NEP models were trained concurrently to quantify uncertainty. Configurations with high force prediction variance were extracted via farthest point sampling (FPS) and then fed into single-point DFT calculations for model refinement. This iterative cycle was repeated over 25 times until the model demonstrated robust stability (no outliers with high prediction uncertainty were detected) within the target thermodynamic configuration space (simulation temperature $T$=300–1000 K, pressure $P$<1000 bar). Comprehensive details regarding the training protocol are provided in the **Methods** section.

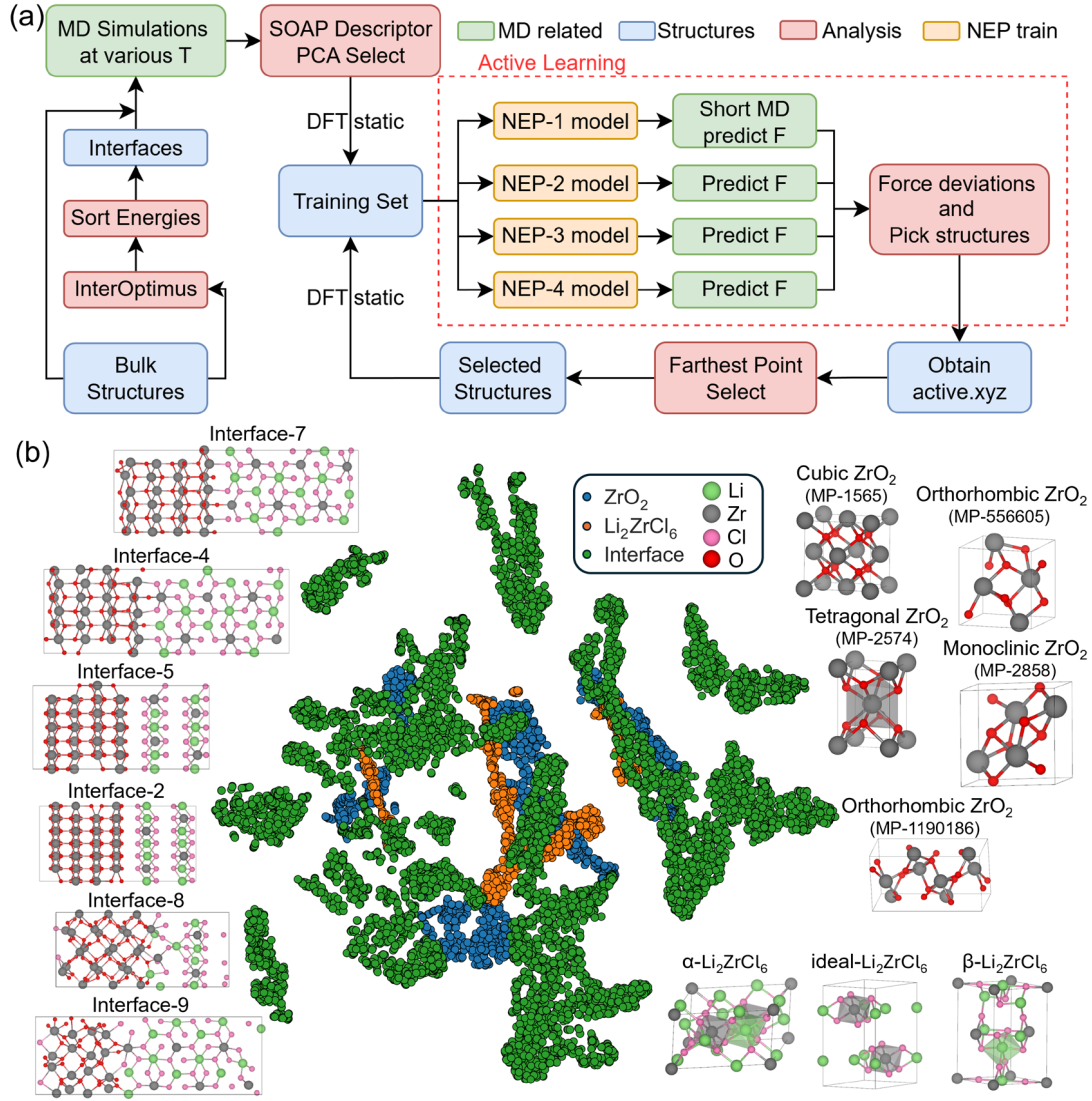

**Figure 1.** (a) The schematic diagram of the general NEP training and active learning workflow. Bulk and representative interface structures are selected for MD simulations, whose trajectories are carefully filtered and calculated by DFT to establish the initial training set. An active learning protocol is later adopted for further refinement of the NEP model. More details are given in the Methods session. (b) Sketch-map visualization of the dataset for the $Li_2ZrCl_6$-$ZrO_2$ interfacial system. In the middle part, a t-SNE visualization of local atomic environments encoded by NEP descriptors reveals the bulk $ZrO_2$ environments in blue, bulk $Li_2ZrCl_6$ environments in orange and interface environments in green. The structures on the upper right, lower right and left provide examples of the representative atomic environments. The detailed interface structure compositions were listed in **Table S1**.

The training dataset comprehensively covers both the bulk phases of $ZrO_2$ and $Li_2ZrCl_6$ and the $Li_2ZrCl_6$-$ZrO_2$ interfacial systems, as elucidated by the t-distributed stochastic neighbor embedding (t-SNE) projection in **Figure 1b**. For the bulk $Li_2ZrCl_6$, three distinct polymorphs have been considered (**Figure S1a**): (1) $\alpha$-$Li_2ZrCl_6$ ($P\bar{3}m1$), which is isostructural to $Li_3YCl_6$; (2) $\beta$-$Li_2ZrCl_6$ ($C2/m$), characterized by alternating $Li^+$ and $Zr$ layers; and (3) the layered phase, denoted as ideal-$Li_2ZrCl_6$ ($P\bar{3}1c$).[23, 26, 29] DFT calculations identify ideal-$Li_2ZrCl_6$ as the ground state (see **Methods** section), with the $\alpha$- and $\beta$-$Li_2ZrCl_6$ phases being 11 and 24 meV/atom higher than the ideal-$Li_2ZrCl_6$ phase. Regarding the $ZrO_2$ component, the monoclinic polymorph (MP-2858, $P2_1/c$)[38] remains stable up to 1400 K before a transition to tetragonal symmetry (MP-2574, $P4_2/nmc$)[39] occurs.[40] Although the cubic phase (MP-1565, $Fm\bar{3}m$)[41] is thermodynamically favored only above 2600 K,[42] both the cubic and tetragonal high-temperature phase can be obtained at lower temperatures by doping zirconia with yttrium.[43] To ensure training set diversity across all the possible configurational spaces, two additional orthorhombic polymorphs, $Pca2_1$ (MP-556605)[44] and $Pbca$ (MP-1190186)[45], were also integrated into the dataset (**Figure S1b**).

For the interfacial systems, we focused on heterostructures composed of all five types of $ZrO_2$ paired with the $\alpha$- and ideal-$Li_2ZrCl_6$. Interface candidates were generated using the InterOptimus package,[46] which systematically samples Miller indices and surface terminations subject to predefined lattice matching criteria. Subsequently, these structures underwent geometry optimization utilizing the MACE-MPA-0 foundation model and were ranked energetically within each interface subclass. The two lowest-energy configurations for each type were identified as thermodynamically plausible interfaces and selected as seed structures to initialize the initial training set generation and active learning (**Table S1**). Detailed generation protocols are provided in the **Methods** Section.

To assess the overall accuracy of the trained NEP model, we computed the root-mean-square errors (RMSEs) for energy, force, and virial components across the entire training dataset, yielding values of $\varepsilon_E^{RMSE}$ =5.6 meV/atom, $\varepsilon_F^{RMSE}$ =219.3 meV/Å and $\varepsilon_v^{RMSE}$ =17.4 meV/atom, respectively. **Figure 2a-c** displays the parity plots comparing NEP predictions with DFT data, demonstrating a high degree of correlation and confirming the model's capability to accurately reproduce the reference PES. Beyond statistical metrics, the physical predictive power of the model was validated against key material properties. Specifically, the model successfully reproduced the nudged elastic band (NEB) migration barriers for $\alpha$- and ideal-Li$_2$ZrCl$_6$ along selected pathways (**Figure S2**) and the energy-volume curves for all bulk polymorphs (**Figure S3**), capturing general energetic trends with minimal deviation. Furthermore, the generalization capability was rigorously tested on a hold-out set comprising 270 interface configurations generated via InterOptimus and relaxed by ASE with MACE-MPA-0 model, which were not included in the training set. As shown in **Figure S4**, the comparison reveals a low energy discrepancy of 10.0 meV/atom between NEP and DFT predictions, underscoring the robustness and transferability of the developed NEP model.

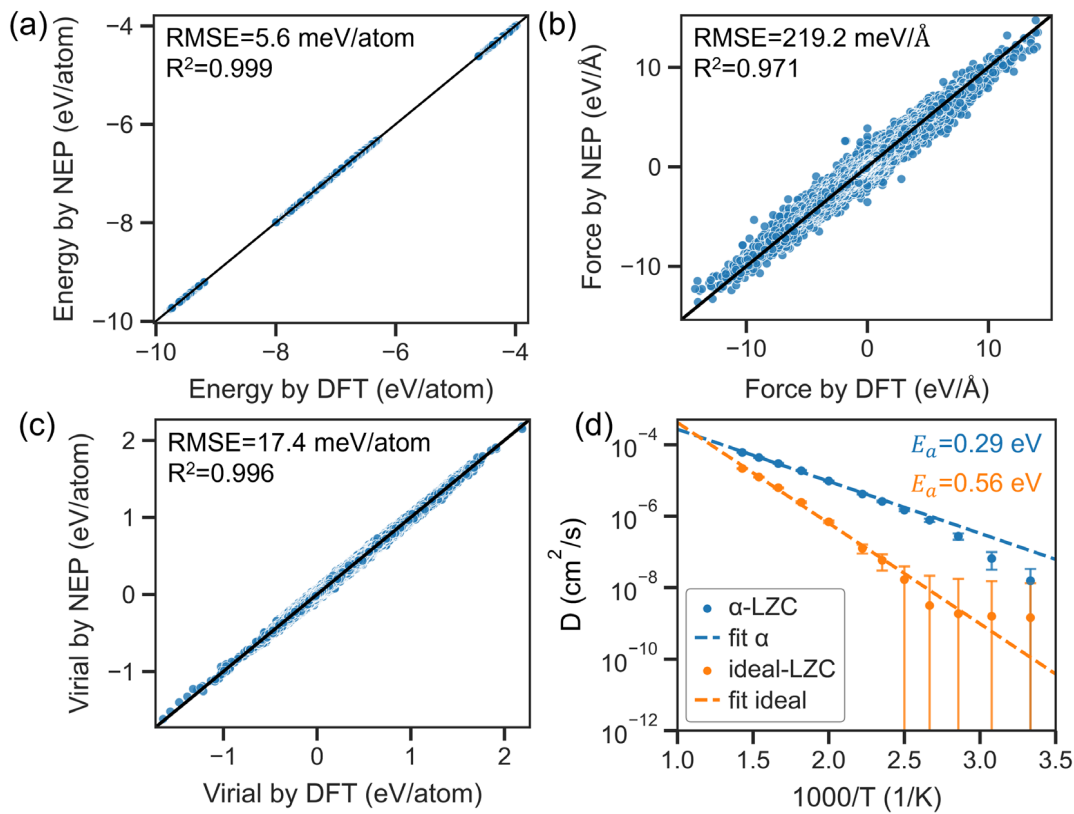

**Figure 2.** RMSE of the (a) energies, (b) forces and (c) virials predicted by NEP compared to DFT on the training dataset. (d) The calculated diffusion coefficients of $\alpha$-$Li_2ZrCl_6$ and ideal-$Li_2ZrCl_6$ were plotted with respect to the inverse of $T$. And the $E_a$ linear fitting of data points from 450 K to 700 K was shown in dash lines.

Having established the accuracy of the NEP model in energy predictions, we proceeded to evaluate its performance in describing dynamic properties, specifically $Li^+$ transport kinetics. To rigorously quantify the distinct diffusion behaviors between the $ZrO_2$-$Li_2ZrCl_6$ composite and the pristine electrolyte, it is imperative to first benchmark the model against experimental measurements of the ionic conductivity for bulk $Li_2ZrCl_6$. This validation ensures that the calculated diffusivity serves as a reliable baseline for subsequent interfacial analysis. Trends in Li$^+$ diffusivity and mean square displacement (*MSD*) as a function of temperature for *α*- and ideal-Li$_2$ZrCl$_6$ were presented in **Figure S5a, b**. Arrhenius plots derived from simulation data over the *T*=450–700 K range yield the $E_a$ presented in **Figure 2d**. The calculated room-temperature ionic conductivity ($\sigma_{RT}$) and $E_a$ for crystalline *α*-Li$_2$ZrCl$_6$ is 0.803 mS/cm and 0.29 eV, demonstrating excellent agreement with experimental benchmarks (0.29~0.81 mS/cm for *α*-Li$_2$ZrCl$_6$).[23, 24, 26, 28, 47] In contrast, the crystalline ideal-Li$_2$ZrCl$_6$ exhibits a 0.20 eV higher $E_a$ and a significantly lower $\sigma_{RT}$ (0.066 mS/cm), which is an order of magnitude below reported experimental values.[29] This result is physically consistent with the limited mobility expected in fully occupied layer of LiCl$_6$-ZrCl$_6$ in ideal-Li$_2$ZrCl$_6$. The experimentally observed enhanced conductivity in ideal-Li$_2$ZrCl$_6$ likely comes from the special ball-milling procedure for preparation of the sample that leads to fine-grained polycrystalline and a high degree of amorphization[48], which deserve thorough investigations in future studies. Notably, the MLMD simulations at 300 K reveal a spontaneous structural reorganization of *β*-Li$_2$ZrCl$_6$ into a pseudo-layered morphology within several hundred picoseconds. As illustrated in **Figure S5**, this transformation involves an interlayer glide of approximately 2.2 Å. Following this relaxation, the ionic conductivity converges with that of the ideal-Li$_2$ZrCl$_6$, underscoring the intrinsic dynamical similarity of these layered frameworks irrespective of relative position of neighboring layers.

## 2.2. The cubic $ZrO_2$ (111)/ideal-$Li_2ZrCl_6$ (001) interface

Among the considered polymorphs, the cubic $ZrO_2$ (111) surface is the sole candidate capable of forming a heterointerface with the ideal-$Li_2ZrCl_6$ (001) plane while maintaining a lattice mismatch strain below 1%. As depicted in **Figure 3a**, the $ZrO_2$ (111) surface terminates in undercoordinated $ZrO_7$ polyhedrons, where the O is not bonded to enough cations. This generates a local space-charge layer that exerts a strong electrostatic attraction on $Li^+$. Conversely, the ideal-$Li_2ZrCl_6$ (001) surface comprises thermodynamically stable, fully occupied $LiCl_6$-$ZrCl_6$ layers without broken bonds. To investigate the dynamic evolution of this interface, we performed 20 ns MLMD simulations at 300, 400, and 500 K under the NPT ensemble with constant number, pressure and temperature (see details of the MLMD simulations in the **Methods** section). Our structural analysis focuses on the 400 K trajectory, a temperature chosen to expedite $Li^+$ diffusion and interfacial equilibration while preserving the bulk integrity of the electrolyte. As expected, the interfacial system (final snapshot at 400 K shown in **Figure 3b**) reveals severe spontaneous reconstruction involving the nearest ideal-$Li_2ZrCl_6$ layer. To quantify this structural evolution, we defined a cutoff radius of 3.5 Å for the $Li^+$ coordination environment based on the radial distribution function (RDF) of bulk $Li_2ZrCl_6$ (**Figure S7a**). **Figure 3c** presents the averaged coordination analysis for the last 1 ns trajectory of MLMD simulation at 400 K, categorizing $Li^+$ environments by coordination number (*CN*) along the surface-normal (*z*) axis. As determined by the *CN* values, the interfacial region is dominated by undercoordinated

Li$^+$ species (Li$^+$ *CN*< 6), indicative of significant amorphization driven by the space-charge effects of the cleaved ZrO$_2$ surface. In contrast, the bulk-like interior retains the characteristic six-fold coordination (Li-6Cl) of the pristine ideal-Li$_2$ZrCl$_6$ phase (**Figure S7c**). Further insights into the spatiotemporal evolution of Li$^+$ distribution are provided in **Figure 3d** (Larger figure with numbers given in **Figure S8**). The simulation initiates with a well-ordered layered structure containing 196 Li$^+$ per plane. Over time, Li$^+$ in the nearest layer to the interface disperse across a ~8 Å range along the *z*-direction, accompanied by Li$^+$ migration from the second-nearest layer. This interfacial reconstruction persists for approximately 2 ns, resulting in a disordered, amorphized region that locally disrupts the symmetry of the ideal-Li$_2$ZrCl$_6$ lattice.

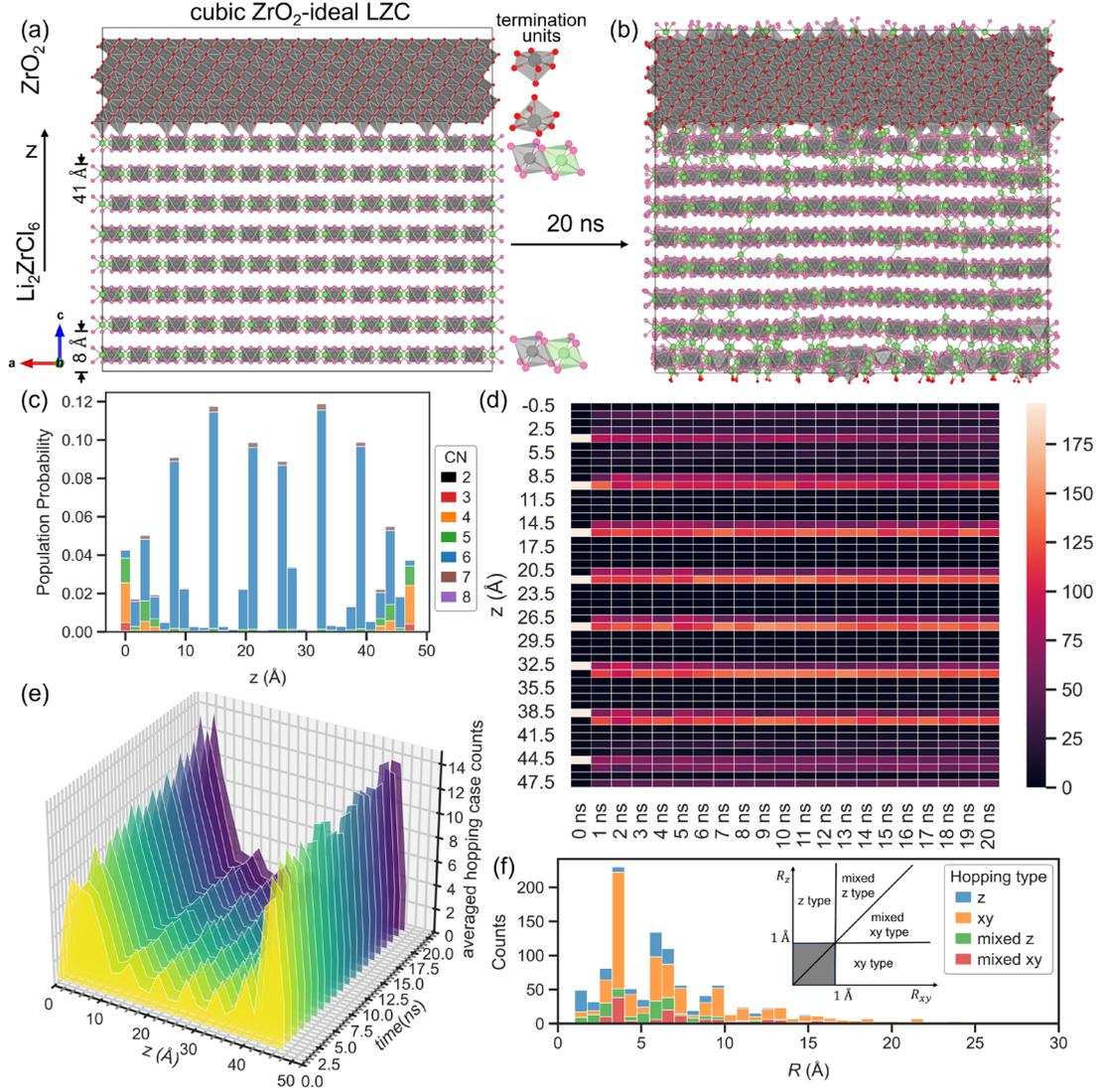

**Figure 3.** The structure of MP-1565 $Fm\bar{3}m$ $Zr_{1728}O_{3456}$ (111)/ $P\bar{3}1c$ ideal-$Li_{1568}Zr_{784}Cl_{4704}$ (001) interface structure for the (a) initial (t=0 ns) and (b) final frame (t=20 ns). The maximum length strain for this combination is 0.5%. The interface region ($z<8$ Å and $z>41$ Å) is marked in the figure. (c) The Li$^+$ coordination analysis (Li-O and Li-Cl) versus the Li$^+$ z-axis coordinates averaged for the last 1 ns of the 20 ns NPT cubic $ZrO_2$ (111)/ideal-$Li_2ZrCl_6$ (001) MLMD simulation at 400 K. (d) The number distribution of Li$^+$ along z axis with respect to the simulation time. (e) The averaged number of hopping event distributions along the z axis with respect to the simulation time. (f) The Li$^+$ diffusion directions can be categorized into four different types by comparing projected displacements on the $xy$-plane ($R_{xy}$) and along the $z$-axis ($R_z$) as shown in the inner panel. The distribution of the overall displacements (after simulation time $t$=2 ns) for each Li$^+$ were plotted, where dominant xy-plane diffusion can be observed.

To characterize the interfacial dynamics, we first established a criterion for identifying Li$^+$ hopping events. A hop is defined to occur when the displacement of a Li$^+$ between consecutive sampled frames (every 10 ps) exceeds 1.25 Å, a threshold corresponding to approximately half the typical Li–Cl bond length. The frequency of such events, averaging over a 1 ns trajectory (100 frames, starting from $t = 1$ns), is plotted as a function of the z-coordinate in **Figure 3e**. The disordered interface regions, characterized by undercoordinated Li$^+$ species (*CN*<6), exhibit a markedly elevated hopping frequency, whereas the crystalline bulk interior remains dynamically inactive. Given the negligible fluctuation in Li$^+$ population within the bulk region noted previously, these hopping events are inferred to be predominantly local, lacking long-range transport along the surface-normal direction. To further decouple the directional components of Li$^+$ transport, we analyzed the displacement $R$ for each Li$^+$ after t=2 ns (excluding the initial equilibration phase). Based on the projected displacements on the $xy$-plane ($R_{xy}$) and along the $z$-axis ($R_z$), Li$^+$ trajectories were classified into four distinct modes (**Figure 3f**): (1) $z$ type hopping ($R_{xy} < 1$ Å, $D_z \geq 1$ Å); (2) $xy$ type hopping ($R_{xy} \geq 1$ Å, $R_z < 1$ Å); (3) mixed $z$ type hopping ($R_z > R_{xy} \geq 1$ Å); (4) mixed $xy$ type hopping ($R_{xy} > R_z \geq 1$ Å). The displacement distribution reveals that motion along the $z$-axis (pure and mixed) is confined to a range of <15 Å, indicating a spatially restricted diffusion pathway originating from the bulk and terminating at the interface. In sharp contrast, transport in the $xy$-plane (pure and mixed) is uninhibited by the ZrO$_2$ boundary, with displacements extending beyond 20 Å, owing to the fast 2D diffusion channels parallel to the interface.

## 2.3. The tetragonal ZrO$_2$ (101)/ideal-Li$_2$ZrCl$_6$ (101) interface

The tetragonal ZrO$_2$ (101) surface forms a coherent interface with the ideal-Li$_2$ZrCl$_6$ (101) plane, characterized by a minimal lattice mismatch strain of <1% (**Figure S9a**). Structurally, unlike the (001) orientation that presents alternating cation/anion layers orthogonal to the $z$-axis, the ideal-Li$_2$ZrCl$_6$ (101) surface possesses a z-direction diffusion tunnel. The interface features a termination of LiCl$_5$ and ZrCl$_4$ units at the upper boundary, and LiCl$_4$ and ZrCl$_3$ at the lower boundary. Applying the same analytical protocol, we observe significant interfacial reconstruction accompanied by a proliferation of undercoordinated Li$^+$ species (*CN*<6), as detailed in **Figure S10a**. However, a distinct contrast to the (001) case emerges: the bulk integrity of the ideal-Li$_2$ZrCl$_6$ (101) electrolyte is largely preserved. The Li$^+$ number density map (**Figure S10b**) exhibits a well-ordered periodic pattern, where the Li$^+$ deficient regions (dark lines) correlate precisely with the Zr-rich layers as shown in **Figure S9b**, indicating that the reconstruction is confined to the immediate interface. Dynamical analysis (**Figure S10c**) reveals that Li$^+$ hopping remains suppressed within the bulk region. This confirms that the kinetic inertness of the fully occupied LiCl$_6$-ZrCl$_6$ framework is an intrinsic property, independent of crystallographic orientation. Conversely, the interfacial regions remain the primary contributor of Li$^+$ hopping activity. The distribution of net Li$^+$ displacements after simulation time t=2 ns indicates an enhanced probability of $z$-directional diffusion in the <5 Å range; nevertheless, long-range transport (>15 Å) remains rare. These findings suggest that simply altering the interfacial orientation is insufficient to overcome the intrinsic geometric barriers to $z$-directional transport in the absence of external driving forces.

## 2.4. The $ZrO_2$/$α$-$Li_2ZrCl_6$ interfaces

As for other $Li_2ZrCl_6$ phases with intrinsic sublattice disorder, we identified two low-strain (<1%) heterostructures involving $α$-$Li_2ZrCl_6$ and $ZrO_2$ polymorphs: (1) cubic $ZrO_2$ ($22\bar{1}$)/$α$-$Li_2ZrCl_6$ (001) and (2) monoclinic $ZrO_2$ (010)/$α$-$Li_2ZrCl_6$ (201). The structural configuration of the first interface is depicted in **Figure S11a**. The upper boundary features $ZrO_4$ and $ZrO_5$ polyhedrons interfacing with $LiCl_6$ and $ZrCl_6$ octahedrons, while the lower boundary involves $ZrO_5$ and $ZrO_8$ units contacting undercoordinated $LiCl_3$ and $ZrCl_3$ units. Post-simulation analysis of the MLMD simulations at 400 K (**Figure S11b**) reveals distinct reconstruction behaviors. At the upper interface, while $ZrCl_6$ units remain largely intact (but distorted), the $LiCl_6$ octahedrons undergo severe disintegration: their $Li^+$ ions migrate toward surface oxygen sites driven by electrostatic attraction, while the Cl ions were captured by the Zr. This results in an overall loosely connected interface sustained primarily by $ZrCl_6$-$LiCl_xO_y$ bonding. On the other side, the lower interface develops mixed-anion $ZrO_3Cl_3$ and $LiO_3Cl_3$ coordination environments, indicating stronger interfacial adhesion. A striking anisotropy is observed: while $Li^+$ is significantly depleted at the upper $Li_2ZrCl_6$ interface, the lower interface retains a high Li$^+$ concentration. This behavior stems from the intrinsic sublattice dynamics of α-Li$_2$ZrCl$_6$. As shown in **Figure S12**, Li$^+$ ions exhibit a preferential distribution between alternating layers, establishing a Li$^+$ population ratio of approximately $N_{rich}:N_{poor} \approx 2:1$ between the Li$^+$-rich and Li$^+$-poor layers. The upper interface coincides with a Li-poor layer; consequently, the local Li$^+$ inventory is rapidly exhausted by surface charge compensation without sufficient replenishment from the bulk. This depletion is quantified in **Figure 4a**, where the Li$^+$ population adsorbed at the upper ZrO$_2$ surface is physically decoupled from the electrolyte bulk by a 3~4 Å gap. In contrast, the Li$^+$ network at the lower interface maintains continuous connectivity with the electrolyte. The bulk region retains the structural signature of pristine α-Li$_2$ZrCl$_6$ (**Figure S7b**). Regarding the second heterostructure (**Figure S11c**), the monoclinic ZrO$_2$ (010) surface feature ZrO$_4$ units contacting a complex mixture of LiCl$_x$ and ZrCl$_x$ (x=3, 4, 5) on both sides. The equilibrated structure (last frame) at 400 K (**Figure S11d**) exhibits substantial local disorder. Coordination analysis (**Figure 4b**) confirms that Li$^+$ at the interface are predominantly undercoordinated ($CN$<6), reflecting a degree of amorphization and structural perturbation more severe than that observed in the previous heterostructure (cubic ZrO$_2$ ($22\bar{1}$)/α-Li$_2$ZrCl$_6$ (001)).

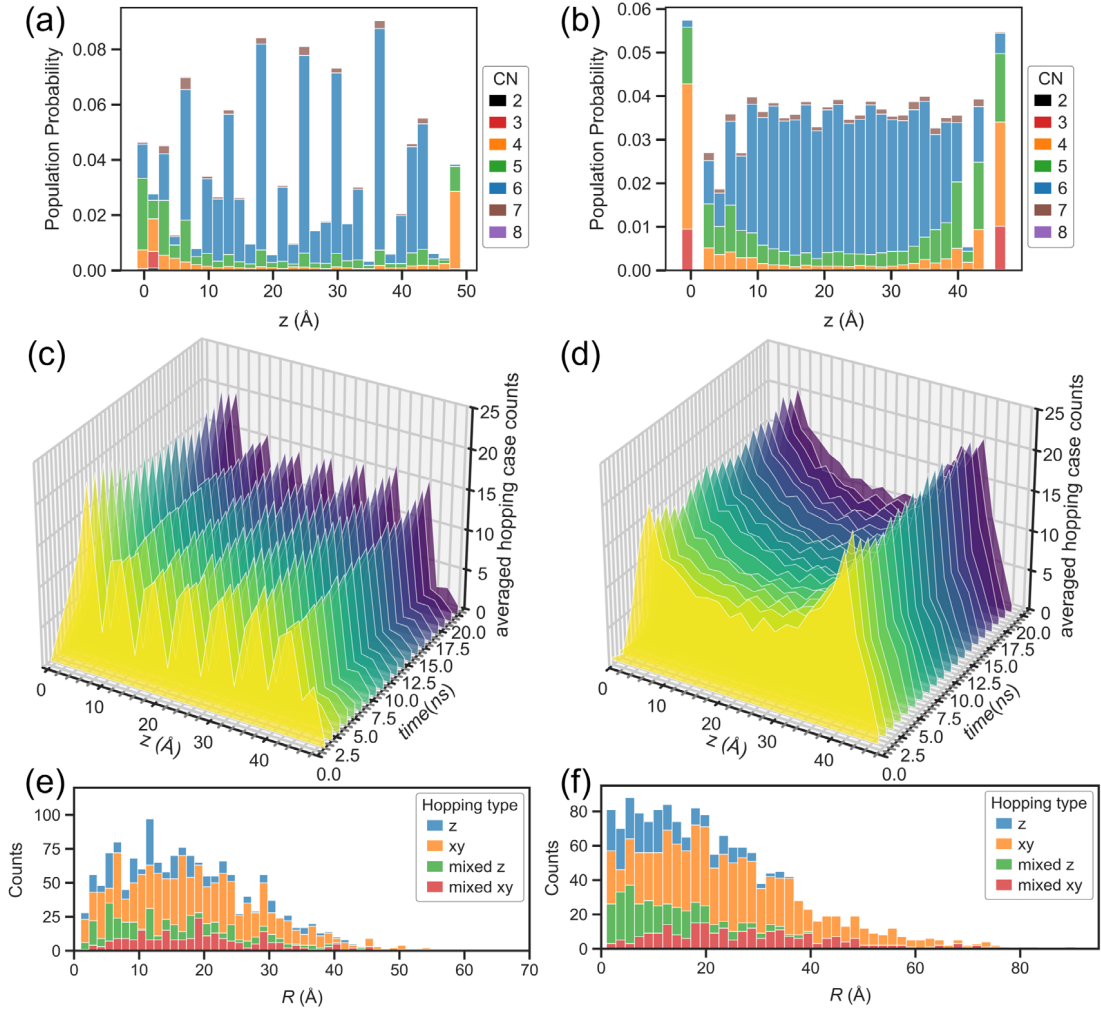

**Figure 4.** The structural and dynamical analysis were conducted for (a, c, e) MP-1565 $Fm\bar{3}m$ ZrO$_2$ (22$\bar{1}$)/$P\bar{3}m1$ $\alpha$-Li$_2$ZrCl$_6$ (001) interface and (b, d, f) MP-2858 $P2_1/c$ ZrO$_2$ (010)/$P\bar{3}m1$ $\alpha$-Li$_2$ZrCl$_6$ (201) interface. (a, b) The Li$^+$ coordination analysis versus the Li$^+$ z-axis coordinates averaged for the last 1 ns of the 20 ns NPT interface MLMD simulation at 400 K. (c, d) The averaged number of hopping event distributions along the z axis with respect to the simulation time. (e, f) The distribution of the overall displacements (after $t$=2 ns) for each Li$^+$ were plotted.

The dynamical properties for these heterostructures were elucidated using the hopping event analysis presented in **Figure 4c, d**. In stark contrast to the kinetically stable bulk of the ideal-Li$_2$ZrCl$_6$ phase, significant Li$^+$ hopping activity persists throughout the bulk of the $\alpha$-Li$_2$ZrCl$_6$ phase. The hopping profiles exhibit distinct topologies dictated by crystallographic orientation: the $\alpha$-Li$_2$ZrCl$_6$ (001) interface displays a pronounced discrete peak-valley periodicity (**Figure 4c**), whereas the (201) interface manifests a continuous distribution (**Figure 4d**). While interfacial amorphization typically enhances diffusivity, the upper interface of the cubic ZrO$_2$ ($22\bar{1}$)/$\alpha$-Li$_2$ZrCl$_6$ (001) system presents a notable anomaly. Here, the hopping frequency diminishes rather than increases. This suppression is mechanistically ascribed to severe carrier depletion; although structural disorder theoretically lowers migration barriers, the scarcity of mobile Li$^+$ limits overall transport. This underscores a critical insight into the insulator-SSE composite design rule that interfacial reconstruction enhances conductivity only when coupled with an adequate population of mobile charge carriers. The net displacement distributions after t=2 ns (**Figure 4e**, **f**) further quantify this enhanced mobility. Unlike the ideal-Li$_2$ZrCl$_6$ phase, where $z$-directional excursions are spatially confined (<15 Å), Li$^+$ in the $\alpha$-phase readily traverse distances exceeding 20 Å along the $z$-axis. Furthermore, maximum displacements in the interfaces composed of the $\alpha$-Li$_2$ZrCl$_6$ and ZrO$_2$ can easily extend beyond the ~30 Å limit observed in the interface of ideal-Li$_2$ZrCl$_6$. These results confirm that the interfaces with $\alpha$-Li$_2$ZrCl$_6$ offer a geometrically less restricted environment, thereby facilitating superior long-range ion transport.

## 2.5. Coordination environments and local distortions in the interfacial regions

To elucidate the structural origins governing enhanced Li$^+$ diffusion at interfaces, we quantified the local coordination environments using the Continuous Symmetry Measure (CSM) as the structural descriptor. In this framework, a CSM value of 0 denotes a perfectly symmetric coordination geometry, while a maximum value of 100 corresponds to an infinitely elongated configuration along a single axis (details in SI, **Figure S13**).[49-51] **Figure S14** correlates the Li$^+$ coordination polyhedral volume with CSM values for structures extracted from the final snapshot of the 400 K NPT simulation. Analysis of the bulk phases (**Figure S14a, b**) reveals a highly localized distribution that distortions are minimal (CSM<3) and polyhedral volumes cluster in the 20-25 Å$^3$ range, which is characteristic of LiCl$_6$ octahedrons. Distinct clusters at lower volumes correspond to undercoordinated LiCl$_5$ square pyramids (10~13 Å$^3$) and LiCl$_4$ units (5~8 Å$^3$). Notably, bulk $\alpha$-Li$_2$ZrCl$_6$ features a significantly higher population of these undercoordinated polyhedrons compared to the ideal phase, which retains a concentrated, highly ordered octahedron distribution. This corroborates previous findings regarding the critical role of heavy-cation-induced structural disorder.[25] In the heterostructures (**Figure S14c-f**), the population density maps (visualized via hexagonal binning) expand to occupy a broad region of the CSM-volume phase space. To decouple interfacial effects, we spatially segregated Li$^+$ into interfacial populations (within 8 Å of the interface, **Figure 5** right panels) and bulk-like populations (**Figure 5** left panels). The interfacial $Li^+$ ions exhibit significant dispersion in the CSM-volume plane, characterized by high geometric distortion and reduced polyhedral volumes. This structural signature is indicative of local amorphization and the prevalence of undercoordinated species at the interface. Conversely, the non-interfacial $Li^+$ ions retain a coordination environment virtually identical to that of the pristine bulk phases.

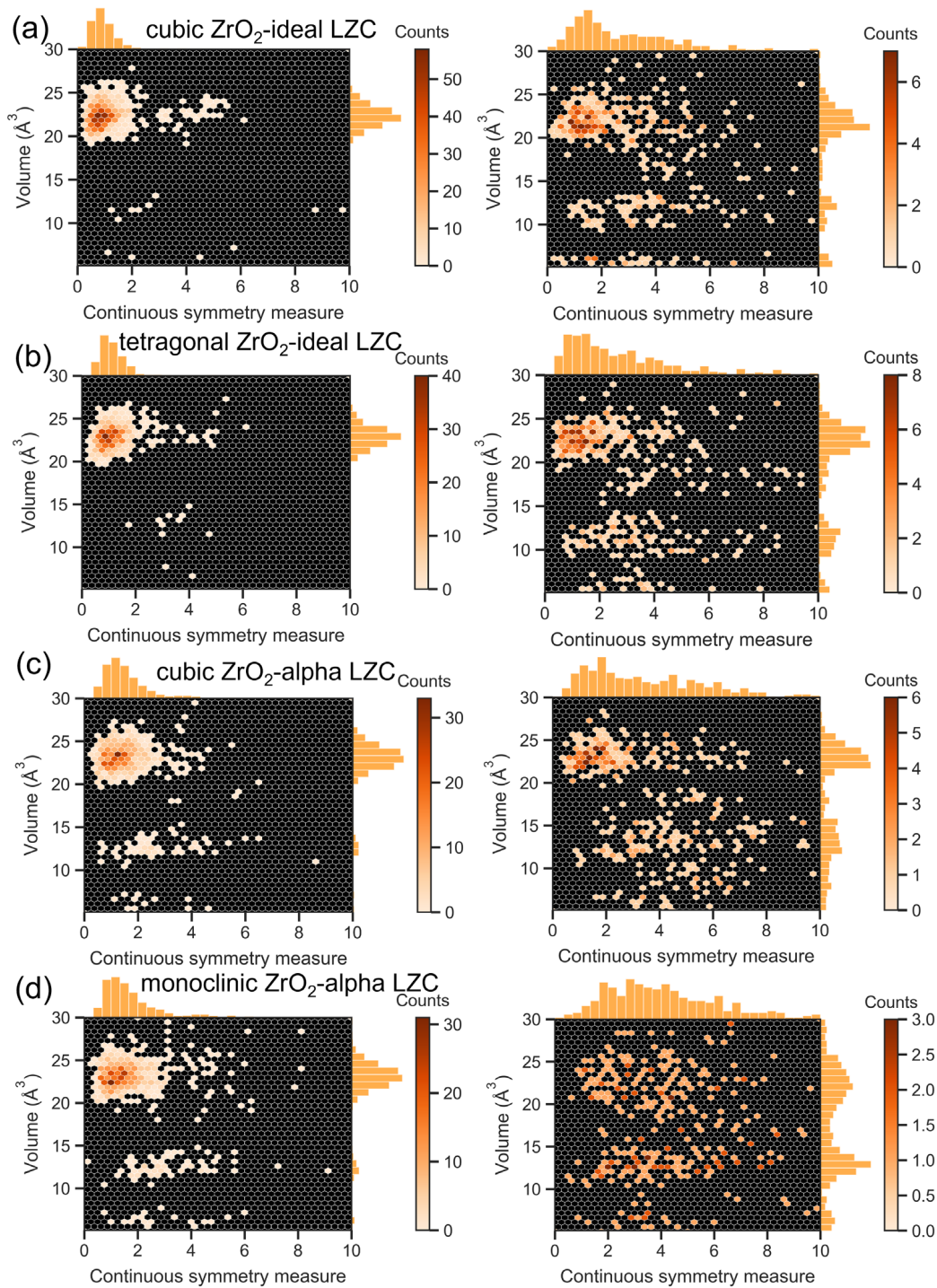

**Figure 5.** The (left panel) bulk and (right panel) interface region Li$^+$ environment for (a) cubic ZrO$_2$ (111)/ideal-Li$_2$ZrCl$_6$ (001), (b) tetragonal ZrO$_2$ (101)/ideal-Li$_2$ZrCl$_6$ (101), (c) cubic ZrO$_2$ (22$\bar{1}$)/α-Li$_2$ZrCl$_6$ (001), and (d) monoclinic ZrO$_2$ (010)/α-Li$_2$ZrCl$_6$ (201). The volume of polyhedron formed by Li$^+$ coordination is plotted against Continuous Symmetry Measure (CSM) in a hexagonal histogram. The deeper the color, the greater number of entries populated in a hexagon. The black color means no count in that hexagon.

The enhanced dynamics observed at under-coordinated and geometrically distorted Li$^+$ sites corroborate the observation of Li$^+$ diffusion promotion in amorphous SSEs.[52] This correlation is further substantiated by the significantly higher Li$^+$ hopping ratios recorded in the interfacial regions relative to the bulk domains. As elucidated by Qiao et al.,[53] the intrinsic structural disorder within amorphous electrolytes creates continuous percolating pathways for ionic migration, thereby allowing ionic conductivities to approach theoretical limits via unique conduction mechanisms. Consequently, the incorporation of ZrO$_2$ into the Li$_2$ZrCl$_6$ matrix effectively engenders a network of fast-diffusion channels. This enhancement, however, presents a thermodynamic trade-off, as a fraction of the Li$^+$ inventory is immobilized at the heterointerface to compensate for the space-charge layer.

## 3. Conclusions

In summary, we developed a robust MLFF within the NEP framework to bridge the gap between *ab initio* accuracy and the spatiotemporal scales required to model bulk Li$_2$ZrCl$_6$ and its heterostructures with ZrO$_2$. By leveraging large-scale, nanosecond-level MD simulations, we demonstrated the atomistic origins of the experimentally observed conductivity enhancement at the ZrO$_2$/Li$_2$ZrCl$_6$ interface. Our results reveal that interfacial formation induces spontaneous structural reconstruction driven by space-charge compensation. This process generates localized amorphous domains enriched with undercoordinated ($CN$<6) and geometrically distorted Li$^+$ sites, as quantified by high CSM values. These disordered regions function as high-speed percolation pathways, exhibiting Li$^+$ hopping rates significantly superior to those of the crystalline bulk. The interfacial Li$^+$ conduction enhancement is contingent upon the availability of mobile charge carriers. In scenarios where the local Li$^+$ inventory is depleted by surface charge compensation without adequate replenishment from the bulk, the transport benefits of amorphization are effectively negated. Furthermore, our analysis confirms the superior intrinsic diffusivity of $α$-Li$_2$ZrCl$_6$ relative to the ideal phase, underscoring the pivotal role of cation-induced structural disorder. Collectively, these findings unveil the complex interplay between favorable structural reconstruction and detrimental carrier depletion, offering critical insights for engineering high-performance solid-solid interfaces in the next-generation solid-state electrolytes.

## 4. Methods

**DFT Calculation Setup.** DFT calculations, including static calculations, geometry optimizations, and NEB calculations, were performed using the Projector Augmented Wave (PAW) method[54] as implemented in the Vienna Ab initio Simulation Package (VASP) (version 6.4.2). The Perdew-Burke-Ernzerhof (PBE) pseudopotential[55] was used for the DFT exchange-correlation functional, and the DFT-D3 method with Becke-Johnson damping function[56] was selected for van der Waals correction.[57] The plane-wave cutoff energy was set to 550 eV with 0.3 Å$^{-1}$ k-point meshes for static calculations and geometry optimizations. The convergence criteria for energy and force were set to $1 \times 10^{-6}$ eV and 0.01 eV Å$^{-1}$, respectively.

**Bulk Structure Selection.** Five experimentally validated $ZrO_2$ polymorphs with an energy above hull of < 0.1 eV/atom were screened from the Materials Project database[58]: cubic (MP-1565, $Fm\bar{3}m$)[41], tetragonal (MP-2574, $P4_2/nmc$)[39], monoclinic (MP-2858, $P2_1/c$)[38], and two orthorhombic phases (MP-556605, $Pca2_1$;[44] MP-1190186, $Pbca$;[45]). Parallelly, three major $Li_2ZrCl_6$ bulk configurations were evaluated: α-$Li_2ZrCl_6$ ($P\bar{3}m1$), β-$Li_2ZrCl_6$ ($C2/m$) and ideal-$Li_2ZrCl_6$ ($P\bar{3}1c$).[23, 26, 29] To resolve the fractional site occupancy inherent to the α- and β-$Li_2ZrCl_6$, we enumerated possible ordered configurations and selected the lowest-energy structures following geometric relaxation. Energetic analysis revealed that the β-$Li_2ZrCl_6$ phase lies 24 meV/atom above the ground state, and thus the β-phase was excluded from the dataset. Subsequent training and interface construction therefore focused exclusively on the ground-state ideal-$Li_2ZrCl_6$ and α-$Li_2ZrCl_6$ (11 meV/atom above the ground state).

**Interface Structure Selection.** Initial heterostructures were generated using the InterOptimus package[46] by pairing the five selected bulk $ZrO_2$ structures with the ideal- and α-$Li_2ZrCl_6$ phases. To construct the initial training dataset, candidate interfaces were assembled based on diverse crystallographic orientation relationships and surface terminations. We applied geometric constraints limiting the interface area to ≤ 120 Å$^2$ and maximum length strain ≤ 10%. This relatively high strain tolerance was deliberately adopted to ensure computationally affordable system sizes for subsequent DFT calculations. All generated assemblies underwent geometry optimization using the MACE-MPA-0 foundation model[37] via the ASE interface[59]. For each unique orientation relationship, the two lowest-energy configurations were retained, yielding a pool of 36 candidate structures. From this pool, a subset of ten interfaces (spanning all $ZrO_2$/$Li_2ZrCl_6$ combinations) was randomly selected to initialize the NEP active learning cycle. For production-level GPUMD simulations, where minimizing artificial strain is critical, we imposed stricter criteria. The maximum lattice mismatch strain was restricted to <1%, while the allowable interface area was expanded to ≤ 200 Å² to accommodate larger commensurate supercells. This rigorous screening isolated four representative interface systems for detailed dynamic analysis:

(1) MP-1565 $Fm\bar{3}m$ $ZrO_2$ (111)/$P\bar{3}1c$ ideal-$Li_2ZrCl_6$ (001)

(2) MP-2574 $P4_2/nmc$ $ZrO_2$ (101)/$P\bar{3}1c$ ideal-$Li_2ZrCl_6$ (101)

(3) MP-1565 $Fm\bar{3}m$ $ZrO_2$ (22$\bar{1}$)/$P\bar{3}m1$ α-$Li_2ZrCl_6$ (001)

(4) MP-2858 $P2_1/c$ $ZrO_2$ (010)/$P\bar{3}m1$ α-$Li_2ZrCl_6$ (201)

**Initial Training Set.** The foundational training dataset was constructed by combining stochastically perturbed configurations (1-5% strain) with diverse snapshots extracted from MD trajectories of bulk $ZrO_2$, bulk $Li_2ZrCl_6$, and their interface structures. To sample the PES extensively, MD simulations were performed using the MACE-MPA-0 model at $T$ of 300 K, 900 K, 1500 K, and 2000 K. Each simulation spanned 100 ps, with configurations recorded at 0.5 ps intervals. To maximize data efficiency and minimize redundancy, we employed a descriptor-based selection protocol. Atomic environments were first encoded using SOAP descriptors via the DScribe package.[62, 63] Dimensionality reduction was subsequently applied to the descriptor arrays using Principal Component Analysis (PCA). Representative structures were then identified by clustering in the reduced 2D latent space, ensuring broad coverage of the sampled configuration space. Following static DFT calculations on the selected configurations, a total of 4,931 structures were incorporated into the initial training set.

**Molecular Dynamics.** Large-scale dynamical simulations were enabled by the highly efficient NEP implemented within the GPUMD code.[60-64] To ensure robust transferability across diverse regions of the potential energy surface (PES, we employed a rigorous active learning workflow based on the query-by-committee (QBC) strategy[65] (**Figure 1a**). In each iterative cycle, an ensemble of four independent NEP models was trained. To extensively sample the configuration space, including highly distorted structures mimicking those induced by high-energy ball-milling, we performed short (100 ps) MD simulations in both NVT (300–3000 K) and NPT (600–1500 K; $10^2$–$10^4$ bar) ensembles. The maximum force deviation ($\varepsilon$) among the ensemble members served as the uncertainty metric. Configurations exhibiting $\varepsilon$ >0.3 eV/Å were flagged as candidates. To filter out unphysical geometries, structures containing interatomic distances $r_{ij} < 0.7 \times (R_i + R_j)$ (where $R$ denotes the covalent radius[66]) were discarded.

From the remaining pool, a diverse subset was selected *via* FPS, sent for DFT static calculations, and incorporated into the training dataset for the subsequent iteration. The NEP architecture employed radial and angular cutoff radii of 6 Å and 5 Å, respectively. For radial (angular) descriptor components, we used 8 (8) radial functions constructed from a linear combination of 12 (12) basis functions. A single hidden layer containing 50 neurons was employed. To accurately model short-range repulsion and prevent atomic overlap, the Ziegler-Biersack-Littmark (ZBL) potential[67] was activated at distances below 1.2 Å (roughly the O-O bond length in the $O_2$ gas). The model was optimized over $10^5$ generations using the entire training set. The final RMSEs for energies, forces and virials were 5.6 meV/atom, 219.2 meV/Å and 17.4 meV/atom, respectively, where RMSE is defined as

$$RMSE = \sqrt{\frac{1}{N}\sum_{i=1}^{N}(Y_{ML} - Y_{DFT})^2}$$

where Y stands for the energies, forces or virials values.

MLMD simulations were conducted under the NPT ensemble using the GPUMD package. Temperature and pressure (1 bar) were regulated via the Bussi-Donadio-Parrinello (BDP) thermostat[68] and the Stochastic Cell Rescaling (SCR) barostat,[69] respectively. A time step of 1 fs was applied, with center-of-mass linear and angular momenta removed every 10 steps to prevent drift. For bulk $Li_2ZrCl_6$ phases, simulations were performed on large supercells to minimize finite-size effects: $\alpha$-$Li_2ZrCl_6$ (3,888 atoms), $\beta$-$Li_2ZrCl_6$ (2,592 atoms), and ideal-$Li_2ZrCl_6$ (2,592 atoms). The systems were equilibrated at 300 K and subsequently heated to 700 K. Production trajectories spanned 5 ns for $T \leq 500$ K and 2 ns for $T > 500$ K. For interface systems, MLMD NPT simulations were then performed on four interfaces as previously mentioned, (1) MP-1565 $Fm\bar{3}m$ $Zr_{1728}O_{3456}$ (111)/ $P\bar{3}1c$ ideal-$Li_{1568}Zr_{784}Cl_{4704}$ (001), (2) MP-2574 $P4_2/nmc$ $Zr_{1792}O_{3584}$ (101)/$P\bar{3}1c$ ideal-$Li_{1152}Zr_{576}Cl_{3456}$ (101), (3) MP-1565 $Fm\bar{3}m$ $Zr_{1512}O_{3024}$ (22 $\bar{1}$ )/ $P\bar{3}m1$ α-$Li_{1152}Zr_{576}Cl_{3456}$ (001), and (4) MP-2858 $P2_1/c$ $Zr_{1536}O_{3072}$ (010)/$P\bar{3}m1$ α-$Li_{1152}Zr_{576}Cl_{3456}$ (201). These interface models were simulated at 300 K, 400 K, and 500 K. Each production run extended for 20 ns to ensure statistical convergence of transport properties.

**MSD and Diffusion Coefficients.** By heating ideal- and α-phase of bulk $Li_2ZrCl_6$ from 300 K to 700 K, the diffusion coefficient $D$ was calculated using the Einstein relation:

$$D = \frac{1}{2dt} \langle [r(t)]^2 \rangle \tag{1}$$

In the case of three-dimensional diffusion ($d=3$), the $MSD$[70] is given by:

$$MSD = \langle [r(t)]^2 \rangle = \frac{1}{N} \sum_{i=1}^{N} \langle [r_i(t+t_0)]^2 - [r_i(t_0)]^2 \rangle \tag{2}$$

where $r_i(t)$ is the displacement of the $i$-th Li$^+$ at time $t$, the average $\langle ... \rangle$ is applied with respect to $t_0$ (starting time), and $N$ is the number of Li$^+$ in the system. Diffusion coefficients were averaged every 10 datapoints up to 50% of the trajectory (500 frames) and reported along with standard deviations.

**Acknowledgements**

B. Xu, L. Bai, C. Wang and Q. Wu were supported by the Advanced Materials-National Science and Technology Major Project from Ministry of Industry and Information Technology of the People's Republic of China [Grant No. 2024ZD0607100] and the funding from Suzhou Laboratory [Grant No. SK-1502-2024-019(PRE)]. B. Xu and C. Wang gratefully acknowledge additional support from Jiangsu Funding Program for Excellent Postdoctoral Talent. C. Qian gratefully acknowledges additional support from China Postdoctoral Science Foundation [Grant No. 2025M781003].

**Data Availability Statement:**

The training datasets, the NEP model, together with the initial interface structures for MD simulations, are available from figshare upon acceptance of this paper.

**Conflict of Interest Disclosure**

The authors declare no competing financial interests.